\documentclass[twocolumn,showpacs,preprintnumbers,amsmath,amssymb]{revtex4}

\usepackage{epsfig}

\newtheorem{theorem}{Theorem}

\begin{document}


\title{Stabilizer states and Clifford operations for systems of arbitrary dimensions,\\ and modular arithmetic.}

\author{Erik Hostens}
\email{erik.hostens@esat.kuleuven.ac.be}
\affiliation{Katholieke Universiteit Leuven, ESAT-SCD, Belgium}
\author{Jeroen Dehaene}
\affiliation{Katholieke Universiteit Leuven, ESAT-SCD, Belgium}
\author{Bart De Moor}
\affiliation{Katholieke Universiteit Leuven, ESAT-SCD, Belgium}
\date{\today}

\newcommand{\XZ}{X\!\!Z}
\newcommand{\Z}{\mathbb{Z}}
\newcommand{\Vdiag}{{\cal V}_{\mathrm{diag}}}
\newcommand{\Pupps}{{\cal P}_{\mathrm{upps}}}
\newcommand{\Pdiag}{{\cal P}_{\mathrm{diag}}}

\begin{abstract}
We describe generalizations of the Pauli group, the Clifford group and stabilizer states for qudits in a Hilbert space of arbitrary dimension $d$. We examine a link with modular arithmetic, which yields an efficient way of representing the Pauli group and the Clifford group with matrices over $\Z_d$. We further show how a Clifford operation can be efficiently decomposed into one and two-qudit operations. We also focus in detail on standard basis expansions of
stabilizer states.
\end{abstract}


\maketitle

\section{Introduction}
We study stabilizer states and Clifford operations for systems built
from qudits (systems with a $d$-dimensional Hilbert space). We work in
a matrix framework using modular arithmetic, generalizing results for
qubits from Ref.~\cite{D:03}. We put special emphasis on the less
studied case where $d$ is not prime.

The stabilizer formalism has already proved to be useful in many
applications such as quantum error correction, entanglement
distillation and quantum computation \cite{GPhD,DVD:03,Gfault,R:03}.

The $n$-qu\emph{d}it generalized Pauli group and Clifford group and the
related concepts of stabilizer codes and states have been studied in 
various levels of detail in a number of
papers \cite{N:02,V:02,G:98,K:96,AK:01,Gr:03,Gr:04,GrQ,S1,S2}. 

Our motivation is not so much the study of stabilizer codes and their
error correcting capacities, but the study of mathematically
interesting states and operations that could play a role in quantum
algorithms.  Although it is well known that building quantum
algorithms with stabilizer states and Clifford operations only is not
sufficient to disallow efficient simulation on a classical computer,
we think it is likely that the rich structure of this formalism will
play a role in future quantum algorithms. Due to this focus, we pay
attention to describing and realizing Clifford operations in more
detail than is usually needed for coding applications. (To specify a
Clifford operation ''completely'' (that is, up to only a global
phase), one has to specify the image under conjugation of $2n$
independent Pauli operations including the resulting phase, whereas to
realize an encoding operator for a $k$-dimenional code, only $k$
images are needed and the phases are of minor importance.)

Next to presenting known results in an often different, and in our
opinion practical language, we also present results not contained
in the references above.

We give a description of an $n$-qudit Clifford operation by a
$2n\times 2n$ matrix $C$ with entries in $\mathbb{Z}_d$ and a
$2n$-dimensional vector $h$ with entries in $\mathbb{Z}_{2d}$ and
derive necessary and sufficient conditions for $C$ and $h$ to
define a Clifford operation. We give formulas for multiplying and
inverting Clifford operations represented in this way.

We present a decomposition of a general Clifford operation,
specified in full detail by a matrix $C$ and $h$, into a selected set of one and
two-qudit operations, by thinking in terms of matrix manipulations on
$C$ and $h$.

We also focus in detail on the standard basis expansion of stabilizer
states. From Ref.~\cite{GrQ,S1,S2} formulas can be
derived describing the standard basis expansion of graph states by
means of a quadratical form. In Refs.~\cite{S2,GrQ} this is
done for the case when the 1-qudit configuration space
$\{1,\ldots,d\}$ is given the structure of a finite field. In
Ref.~\cite{S1} this space can be any finite Abelian group. In
this paper we consider cyclic groups. Refs.~\cite{S2,GrQ} state the equivalence of graph states and general
stabilizer states. In Ref.~\cite{GrQ} this equivalence is to be
understood as local Clifford equivalence. That is, any $n$-qudit
stabilizer state (with a field as $1$-qudit configuration space) can
be transformed into a graph state through the action of $n$ one-qudit
Clifford operations. In our setting however, as we are not focusing on
codes, we want a description of the original stabilizer state (without
the local Clifford operations) as well. In Ref.~\cite{S2}
another notion of equivalence between graph states and stabilizer
states is used (introducing the concept of auxiliary nodes in the
graph). As a result the standard basis expansion of a general
stabilizer state is not described directly but as a sum of a large
number of states. Moreover, for the case where the configuration space
is not a field (in our case that is when $d$ is not prime) not all
stabilizer states are equivalent to graph states but an extra
condition has to be imposed. In the present paper we work with a more
general description of stabilizer states without this extra condition
(described below by matrices $S$ with possibly more than $n$ columns) and we
give a direct description (without sum) of the standard basis
expansion of general stabilizer states. We believe that standard
basis expansions of stabilizer states can be an essential ingredient
in understanding the action of non-Clifford operations on stabilizer
states.

This paper is structured as follows. Definitions of generalizations of 
the Pauli group and the Clifford group for qudits are given in section~\ref{secPC}, 
together with their matrix representation. Special Clifford operations, that are of 
particular interest in the decomposition of a Clifford operation, are discussed in 
section~\ref{secSC}. An efficient decomposition of a Clifford operation on $n$ qudits 
into a selected set of one and two-qubit Clifford operations, is explained in section~\ref{secDC}. 
In section~\ref{secS}, we define stabilizer states of $n$ qudits and show the expansion in the 
standard basis can be described with linear and quadratic operations.

In the following, by $A=B\mod d$ we mean that all corresponding entries of matrices $A$ and $B$ are equal modulo $d$, where $d$ is an integer different from 0. We will also write $a=b\mod c$ with $a$, $b$ and $c$ vectors, as a shorthand notation for $a_i=b_i\mod c_i$, for every $i=1\ldots n$.

\section{The generalized Pauli group and Clifford group}\label{secPC}
In this section, we discuss the description of the generalized Pauli group on $n$ qudits and the generalized Clifford group in modular arithmetic. Generalizations of the Pauli group to systems of arbitrary dimensions are discussed in Refs.~\cite{N:02,V:02,G:98}. The Clifford group is defined as the group containing all unitary operations that map the Pauli group to itself under conjugation.

\subsection{The generalized Pauli group}
Let $d$ be the Hilbert space dimension of one qudit. We define unitary operations $X^{(d)}$ and $Z^{(d)}$ as follows
\begin{equation} \label{XZ}
\begin{array}{rcl} X^{(d)}|j\rangle & = & |j+1\rangle, \\
                   Z^{(d)}|j\rangle & = & \omega^j|j\rangle, \end{array}
\end{equation}
where $j\in\Z_d$ and $\omega$ is a primitive $d$-th root of unity. Addition in the ket is carried out modulo $d$. Tensor products of these operations will be denoted as follows: for $v,w\in\Z_d^n$ and $a:=\left[\begin{array}{c} v \\ w \end{array}\right] \in\Z_d^{2n}$, we denote
\begin{equation}\label{XZ(a)}
\XZ(a):=X^{v_1}Z^{w_1}\otimes\ldots\otimes X^{v_n}Z^{w_n}.
\end{equation}
From (\ref{XZ}) and (\ref{XZ(a)}), it follows that, for $x\in\Z_d^n$,
\begin{equation}
\XZ(a)|x\rangle=\omega^{w^Tx}|x+v\rangle.
\end{equation}
We define the Pauli group ${\cal P}_n$ on $n$ qudits to contain all $d^{2n}$ tensor products (\ref{XZ(a)}) with an additional complex phase factor $\zeta^\delta$, where $\zeta$ is a square root of $\omega$ and $\delta\in\Z_{2d}$. In the following, we will omit the superscript $(d)$ and refer to the generalized Pauli group simply as Pauli group.

Multiplication of two Pauli group elements can be translated into operations on vectors in $\Z_d^{2n}$ as follows:
\begin{equation} \label{paulimult}
\zeta^{\delta}\XZ(a)\zeta^{\epsilon}\XZ(b)=\zeta^{\delta+\epsilon+2a^TUb}\XZ(a+b),
\end{equation}
where $U:=\left[\begin{array}{cc} 0_n & 0_n \\ I_n & 0_n \end{array}\right]$. Addition in the argument of $\XZ$ is done modulo $d$, and addition in the exponent of $\zeta$ is done modulo $2d$. Eq.~(\ref{paulimult}) yields the commutation relation:
\begin{equation} \label{paulicomm}
\XZ(a)\XZ(b)=\omega^{a^TPb}\XZ(b)\XZ(a),
\end{equation}
\begin{equation}\mbox{where}\quad P=U-U^T\mod d.\end{equation} 
Note that the order of $\XZ(a)$ divides $d$ unless $d$ is an even number and $a^TUa$ is odd. In the latter case the order is $2d$. Indeed, with (\ref{paulimult}) one can easily verify that $\XZ(a)^d=\zeta^{d(d-1)a^TUa}I$. The introduction of a phase $\zeta^\delta$ rather than $\omega^\delta$ is only necessary when $d$ is even. Simplifications for odd $d$ are considered in Appendix~\ref{appsimp}.

\subsection{The generalized Clifford group}
We now define a generalization of the Clifford group on $n$ qudits in an analogous way as for qubits. A Clifford operation $Q$ is a unitary operation that maps the Pauli group on $n$ qudits to itself under conjugation, or
\begin{equation*}
Q{\cal P}_nQ^\dag={\cal P}_n.
\end{equation*}
Because
$Q \XZ(a)\XZ(b)Q^\dag=(Q \XZ(a)Q^\dag)(Q \XZ(b)Q^\dag)$,
it is sufficient to know the image of a generating set of the Pauli group in order to
know the image of all Pauli group elements. $Q$ is then defined up to a
global phase factor. This can be seen as follows. Suppose that two Clifford operations $Q_1$ and $Q_2$ give rise to the same image for every Pauli group element, or: for every $A\in{\cal P}_n:~Q_1AQ_1^\dag=Q_2AQ_2^\dag$. It follows for every $A$ that $Q_2^\dag Q_1A=AQ_2^\dag Q_1$. The only unitary operations that commute with every single Pauli group element are multiples of the identity \footnote{Indeed, let $U=\sum_{x,y\in\Z_d^n}q_{xy}|x\rangle\langle y|$ be a unitary operation that commutes with every Pauli group element. From $Z(w)U=UZ(w)$, for all $w\in\Z_d^n$, where $Z(w)$ stands for $Z^{w_1}\otimes\ldots\otimes Z^{w_n}$, it follows that $q_{xy}=0$ for all $x\not=y$. Thus $U=\sum_{x\in\Z_d^n}q_x|x\rangle\langle x|$. From $U=X(v)UX(v)^\dag$, for all $v\in\Z_d^n$, where $X(v)$ stands for $X^{v_1}\otimes\ldots\otimes X^{v_n}$, it follows that all $q_x$ are equal.}, which completes the proof. We take the generating set of the Pauli group to be $\XZ(E_k),~k=1,\ldots,2n$, where $E_k$ are the standard basis vectors of $\Z_d^{2n}$. We denote their images under conjugation by $Q$ as $\zeta^{h_k}\XZ(C_k)$. We will assemble the vectors $C_k$ as the columns of a matrix $C\in\Z_d^{2n\times 2n}$ and the scalars $h_k$ in a vector $h\in\Z_{2d}^{2n}$. The image $\zeta^\epsilon \XZ(b)$ of $\zeta^\delta \XZ(a)$ under conjugation by $Q$, where $a$ is an arbitrary vector in $\Z_d^{2n}$, can be found by repeated application of (\ref{paulimult}). This yields
\begin{equation} \label{imagepauli}
\begin{array}{rcl} b & = & Ca \mod d, \\
            \epsilon & = & \delta+\bigl(h-\Vdiag(C^TUC)\bigr)^Ta +\\
                     &   & a^T\bigl(2\Pupps(C^TUC)+\Pdiag(C^TUC)\bigr)a \mod 2d, \end{array}
\end{equation} where $\Vdiag(M)$ is defined as the vector containing the diagonal of $M$, $\Pdiag(M)$ the diagonal matrix with the diagonal of $M$ and $\Pupps(M)$ the strictly upper triangular part of $M$. The Clifford operation $Q$ is (up to a global phase factor) completely defined by $C$ and $h$. Note that the rhs of (\ref{imagepauli}) is calculated modulo $2d$, although it contains matrices over $\Z_d$. It can be verified that every entry modulo $d$ in the expression is multiplied by an even factor.

We can compose two Clifford operations $Q$ and $Q'$, which again yields a Clifford operation $Q''=Q'Q$. To find its corresponding $C''$ and $h''$ we have to find the images under the second operation of the images under the first operation of the standard basis vectors. By using (\ref{imagepauli}), we get
\begin{equation} \label{combiclifford}
\begin{array}{rcl} C'' & = & C'C \mod d, \\
                   h'' & = & h+C^Th'+\Vdiag\Bigl(C^T\bigl(2\Pupps({C'}^TUC')+\\
                       &   & \Pdiag({C'}^TUC')\bigr)C\Bigr)-C^T\Vdiag({C'}^TUC')\\&& \mod 2d. \end{array}
\end{equation}

The inverse $Q^\dag$ of a Clifford operation $Q$ defined by $C$ and $h$ is defined by $C'$ and $h'$, where
\begin{equation}
\begin{array}{rcl} C' & = & C^{-1} \mod d, \\
                    h' & = & -C^{-T}\Biggl(h+\Vdiag\Bigl(C^T\bigl(2\Pupps(C^{-T}UC^{-1})+\\
                       &   & \Pdiag(C^{-T}UC^{-1})\bigr)C\Bigr)-C^T\Vdiag(C^{-T}UC^{-1})\Biggr)\\&& \mod 2d, \end{array}
\end{equation}
which can be verified with (\ref{combiclifford}). $M^{-T}$ is short for $\left(M^{-1}\right)^T$. We will show below that $C^{-1}=-PC^TP\mod d$.

\subsection{Conditions on $C$ and $h$}
Not all $C\in\Z_d^{2n\times 2n}$ and $h\in\Z_{2d}^{2n}$ define a Clifford operation. To see this, consider a Clifford operation $Q$ with corresponding $C$ and $h$. From the commutation relation (\ref{paulicomm}) it follows that $C$ is a symplectic matrix, i.e. $C$ satisfies $C^TPC=P\mod d$. Indeed, we have
\begin{eqnarray*}
\XZ(a) \XZ(b) & = & \omega^{a^TPb} \XZ(b) \XZ(a) \\
Q \XZ(a) Q^{\dag} Q \XZ(b) Q^\dag & = & \omega^{a^TPb} Q \XZ(b) Q^{\dag} Q \XZ(a) Q^{\dag} \\
\XZ(Ca) \XZ(Cb) & = & \omega^{a^TPb} \XZ(Cb) \XZ(Ca),
\end{eqnarray*}
where we omitted global phase factors on the lhs and rhs, as they cancel each other. Also,
\begin{equation*}
\XZ(Ca) \XZ(Cb) = \omega^{a^TC^TPCb} \XZ(Cb) \XZ(Ca).
\end{equation*}
Since this holds for every value of $a$ and $b$, it follows that $C$ is symplectic. Note that the inverse of a symplectic matrix $C$ is simply $C^{-1}=-PC^TP\mod d$. Secondly, $h$ satisfies
\begin{equation}\label{condh}
(d-1)\Vdiag(C^TUC)+h=0\mod 2,
\end{equation}
for $\zeta^{h_k}\XZ(C_k)=Q\XZ(E_k)Q^\dag$ has, like $\XZ(E_k)$, order $d$. With (\ref{paulimult}) we have $\left(\zeta^{h_k}\XZ(C_k)\right)^d=\zeta^{d\bigl((d-1)C_k^TUC_k+h_k\bigr)}I$, and it follows that (\ref{condh}) is satisfied. We will prove below that every symplectic $C$ and $h$ satisfying (\ref{condh}) define a Clifford operation $Q$.

\section{Special Clifford operations}\label{secSC}
In this section we present a number of special Clifford operations and their defining $C$ and $h$. These will be of particular interest for the decomposition of an arbitrary Clifford operation into one and two-qubit Clifford operations.
\begin{itemize}
\item The Pauli group elements $\XZ(a)$ are a special class of the Clifford operations. Note that, like for any Clifford operation,  the global phase factor of a Pauli group element cannot be represented. Considering the images of $\XZ(E_k)$, it can be easily verified that $\XZ(a)$ is defined by
\begin{equation*}
\begin{array}{rccl}
C & = & I & \mod d\\
h & = & -2Pa & \mod 2d.
\end{array}
\end{equation*}
\item A Clifford operation acting on a subset $\alpha
\subset \{1,\ldots,n\}$ of $n$ qudits gives rise to a symplectic matrix on the rows and
columns with indices in $\alpha\cup(\alpha+n)$, embedded in an identity matrix
(that is, $C_{kk}=1\mod d$, for every
$k\not\in\alpha\cup(\alpha+n)$ and $C_{kl}=0\mod d$ if $k\neq l$ and $k$ or
$l$ $\not\in\alpha\cup(\alpha+n)$). Also $h_k=0\mod 2d$ if
$k\not\in\alpha\cup(\alpha+n)$.
\item Any invertible linear transformation of the configuration space $|x\rangle\rightarrow|Tx\rangle$ can be realized by a Clifford operation, with $x\in\Z_d^n$ and $T\in\Z_d^{n\times n}$ an invertible matrix modulo $d$. This operation is defined by
\begin{equation*}
\begin{array}{rccl}
C & = & \left[\begin{array}{cc}T & 0\\0 & T^{-T}\end{array}\right] & \mod d\\
h & = & 0 & \mod 2d.
\end{array}
\end{equation*}
This can be verified by looking at the image of $\XZ(a)$, with an arbitrary $a=\left[\begin{array}{c} v \\ w \end{array}\right] \in\Z_d^{2n}$: $Q\XZ(a)Q^\dag$
\begin{eqnarray*}
& = & \left(\sum_{x\in\Z_d^n}|Tx\rangle\langle x|\right)\left(\sum_{y\in\Z_d^n}\omega^{w^Ty}|y+v\rangle\langle y|\right)\\&&\left(\sum_{z\in\Z_d^n}|z\rangle\langle Tz|\right)\\
& = & \sum_{y\in\Z_d^n}\omega^{w^Ty}|Ty+Tv\rangle\langle Ty|\\
& = & \sum_{y\in\Z_d^n}\omega^{w^TT^{-1}y}|y+Tv\rangle\langle y|\\
& = & \XZ(\left[\begin{array}{c}Tv\\T^{-T}w\end{array}\right]).
\end{eqnarray*}
As $C^TUC=U\mod d$, we see with (\ref{imagepauli}) that $h=0\mod 2d$.
Special cases of this class of Clifford operations are qudit permutations, with $C=\left[\begin{array}{cc}\Pi & 0\\0 & \Pi \end{array}\right]$, where $\Pi$ is a permutation matrix, and the two-qudit SUM gate $|x\rangle|y\rangle\rightarrow|x\rangle|x+y\rangle$ with $x,y\in\Z_d$, with
\begin{equation*}
C=\left[\begin{array}{cccc}
1 & 0 & 0 & 0 \\ 1 & 1 & 0 & 0 \\ 0 & 0 & 1 & -1 \\ 0 & 0 & 0 & 1\end{array}\right]\mod d.
\end{equation*}
Note that this operation is a natural generalization of the two-qubit CNOT gate.
\item The \emph{$d$-dimensional discrete Fourier transform} $|x\rangle\rightarrow\frac{1}{\sqrt{d}}\sum_{k=0}^{d-1}\omega^{kx}|k\rangle$ on one qudit, with $x\in\Z_d$, is defined by $C=\left[\begin{array}{cc}0 & -1\\1 & 0 \end{array}\right]\mod d$ and $h=0\mod 2d$. We verify this in the same way as for the invertible configuration space transformation, now with $a=\left[\begin{array}{c} v \\ w \end{array}\right] \in\Z_d^{2}$: $Q\XZ(a)Q^\dag$
\begin{eqnarray*}
\qquad & = & \left(\frac{1}{\sqrt{d}}\sum_{t,u\in\Z_d}\omega^{tu}|t\rangle\langle u|\right)\left(\sum_{y\in\Z_d}\omega^{wy}|y+v\rangle\langle y|\right)\\&&\left(\frac{1}{\sqrt{d}}\sum_{x,z\in\Z_d}\omega^{-xz}|z\rangle\langle x|\right)\\
& = & \frac{1}{d}\sum_{t,y,x\in\Z_d}\omega^{t(y+v)+wy-xy}|t\rangle\langle x|\\
& = & \frac{1}{d}\sum_{y\in\Z_d}\omega^{(t+w-x)y}\sum_{t,x\in\Z_d}\omega^{tv}|t\rangle\langle x|\\
& = & \sum_{x\in\Z_d}\omega^{(x-w)v}|x-w\rangle\langle x|\\
& = & \omega^{-vw}\XZ(\left[\begin{array}{c}-w\\v\end{array}\right]).
\end{eqnarray*}
As $C^TUC=-U^T\mod d$, we see with (\ref{imagepauli}) that $h=0\mod 2d$.
This operation is the qudit equivalent of the Hadamard gate on one qubit.
\item Analogous to the qubit phase gate, a \emph{phase gate} on one qudit can be defined as $|x\rangle\rightarrow\zeta^{x(x+d)}|x\rangle$, with $x\in\Z_d$. This operation corresponds to $C=\left[\begin{array}{cc}1 & 0\\1 & 1 \end{array}\right]\mod d$ and $h=\left[\begin{array}{c}d+1\\0\end{array}\right]\mod 2d$. Indeed, for all $a=\left[\begin{array}{c} v \\ w \end{array}\right] \in\Z_d^{2}$: $Q\XZ(a)Q^\dag$
\begin{eqnarray*}
\qquad & = & \sum_{y\in\Z_d}\zeta^{2wy+(y+v)(y+v+d)-y(y+d)}|y+v\rangle\langle y|\\
& = & \zeta^{v(v+d)}\sum_{y\in\Z_d}\omega^{(v+w)y}|y+v\rangle\langle y|.
\end{eqnarray*}
As $C^TUC=\left[\begin{array}{cc}1&0\\1&0\end{array}\right]\mod d$, $v(v+d)$ must be equal to $\left(h-\left[\begin{array}{c}1\\0\end{array}\right]\right)^Ta+v^2\mod 2d$ according to (\ref{imagepauli}), which is the case for the given $h$.
\end{itemize}

\section{Decomposition of a Clifford operation in one and two-qudit operations}\label{secDC}
In order to prove that any symplectic matrix $C$ and $h$ satisfying (\ref{condh}) define a Clifford operation, we will expand an arbitrary symplectic $C$ into symplectic elementary row operations that can be realized as Clifford operations on maximally two qudits at the same time. What is more, this decomposition is a worthy candidate as a practical realization of a Clifford operation. The possibility of this kind of decomposition into a selected set of one and two-qudit operations is briefly discussed in Ref.~\cite{G:98}. Our scheme is related to the method of Ref.~\cite{N:02} in which Euclid's algorithm is incorporated in order to generate any one-qudit Clifford operation. 

First, we mention that the main problem is realizing $C$, not $h$, for once a Clifford operation $Q$ defined by $C$ and $h$ is realized, we can realize $Q'$ defined by $C$ and an arbitrary $h'$ satisfying (\ref{condh}) by doing an extra operation $\XZ\left(CP\frac{h'-h}{2}\right)$ on the left or $\XZ\left(P\frac{h'-h}{2}\right)$ on the right of $Q$. Note that as both $h$ and $h'$ satisfy (\ref{condh}), $h'-h$ is even.

We first give an overview of the elementary row operations that we will use to transform an arbitrary symplectic matrix $C$ into the $2n\times 2n$ identity matrix $I$. As $I$ is formed by left multiplication of such elementary row operations on $C$, a decomposition of $C$ then consists of the inverses of these operations in reverse order. Since these operations act on maximally two qudits at the same time, they are defined by a symplectic $4\times 4$ or $2\times 2$-matrix embedded in the identity matrix as explained in the preceding section. In the following, we will only show this part of the operations.

Firstly, we consider some configuration space transformations (of the form $C=\left[\begin{array}{cc}T & 0\\0 & T^{-T}\end{array}\right]$). These operations combine only rows from the same block (we call rows $1\ldots n$ the upper block and rows $n+1\ldots 2n$ the lower block) and have a similar action in both blocks at the same time. For instance, we can switch two rows $i$ and $j$ in the upper block: at the same time, rows $n+i$ and $n+j$ in the lower block are also switched. This operation is defined by
\begin{equation*}
C=\left[\begin{array}{cccc}0 & 1 & 0 & 0\\ 1 & 0 & 0 & 0\\ 0 & 0 & 0 & 1\\ 0 & 0 & 1 & 0\end{array}\right]\mod d.
\end{equation*}
Multiplying a row $i$ with an invertible number $r\in\Z_d$ results in multiplying the corresponding row $n+i$ in the other block by $r^{-1}$. A number $r\in\Z_d$ has an inverse if $r$ and $d$ are coprime, i.e. $\gcd(r,d)=1$. This operation is defined by $C=\left[\begin{array}{cccc}r & 0 \\ 0 & r^{-1} \end{array}\right]$. The last configuration space transformation we consider, is adding one row $i$ multiplied by an arbitrary factor $g\in\Z_d$ to another row $j$. At the same time, row $n+j$, multiplied by $-g$, is added to row $n+i$. This operation is defined by
\begin{equation*}
C=\left[\begin{array}{cccc}1 & 0 & 0 & 0\\ g & 1 & 0 & 0\\ 0 & 0 & 1 & -g\\ 0 & 0 & 0 & 1\end{array}\right]\mod d.
\end{equation*}

Secondly, we will also need operations that combine rows of different blocks. Switching two rows $i$ and $n+i$ can be carried out by the discrete Fourier transform. Recall that this operation is defined by $C=\left[\begin{array}{cc}0 & -1\\1 & 0 \end{array}\right]$. After switching of the rows, row $i$ is multiplied by $-1$. By applying the inverse of the discrete Fourier transform, row $n+i$ instead of row $i$ is multiplied by $-1$. Applying $\sum_{x\in\Z_d}\zeta^{gx(x+d)}|x\rangle\langle x|$ (which is the same as applying the phase gate $g$ times) on the $i$-th qudit, with $g\in\Z_d$, results in the addition of row $i$ multiplied by $g$ to its corresponding row $n+i$, according to
\begin{equation*}
\begin{array}{ccc}
C & = & \left[\begin{array}{cc}1 & 0\\g & 1\end{array}\right]\mod d.
\end{array}
\end{equation*}

We could introduce more row operations that define one or two-qudit Clifford operations, but the ones described so far suffice. Next we give a constructive way of transforming $C$ into the identity matrix $I$. If we are able to transform $C$ into $C'$ by transforming columns $C_1$ and $C_{n+1}$ into the corresponding columns $E_1$ and $E_{n+1}$ of $I$, it follows from the symplecticity of $C'$ that the first and $n+1$-th row of $C'$ are equal to the corresponding rows of $I$. We then have
\begin{equation*}
C'=\left[\begin{array}{c|ccc|c|ccc}1      & 0 & \ldots    & 0 & 0      & 0 & \ldots    & 0 \\ \hline
                                   0      &   &           &   & 0      &   &           &   \\
                                   \vdots &   & C_{(11)}' &   & \vdots &   & C_{(12)}' &   \\
                                   0      &   &           &   & 0      &   &           &   \\ \hline
                                   0      & 0 & \ldots    & 0 & 1      & 0 & \ldots    & 0 \\ \hline
                                   0      &   &           &   & 0      &   &           &   \\
                                   \vdots &   & C_{(21)}' &   & \vdots &   & C_{(22)}' &   \\
                                   0      &   &           &   & 0      &   &           &   \end{array}\right]
\end{equation*}
Leaving the first qudit out, we can continue by transforming the second and $n+2$-th column of $C'$ into the corresponding columns of $I$, and so on. This recursive procedure eventually leads to $I$. Now we only have to show how columns $C_1$ and $C_{n+1}$ are transformed into $E_1$ and $E_{n+1}$. Let us first consider the case where the upper left entry $C_{11}$ has an inverse in $\Z_d$. Multiplying the first row by $C_{11}^{-1}$ changes this entry to 1. Next we add the first row, multiplied by $-C_{k1}$, to row $k$, and this for $k=2\ldots n$, setting the $k$-th entry of $C_1$ to 0. The first column now has the form $[1~0~\cdots~0\ |\ C_{n+1,1}'~C_{n+2,1}~\cdots~C_{2n,1}]^T$. Now we add the first row multiplied by $-C_{n+1,1}'$ to row $n+1$, setting the $n+1$-th entry of $C_1$ to 0. The discrete Fourier transform on the first qudit changes $C_1$ into $[0~0~\cdots~0\ |\ 1~C_{n+2,1}~\cdots~C_{2n,1}]^T$. In the same way as for the upper half of $C_1$, we make zeros below the $n+1$-th position. Note that nothing happens to the upper half, for all entries there are 0. Switching the first (now we use the inverse of the discrete Fourier transform) with the $n+1$-th row again yields $E_1$. We call the matrix made so far $C''$. From the symplecticity of $C''$ it follows that $C_{n+1,n+1}''=1\mod d$, and we can repeat for the $n+1$-th column the same procedure we did for the first column. Note that none of the operations yielding $E_{n+1}$ out of $C_{n+1}''$ will affect $C_1''=E_1$, except the discrete Fourier transform and its inverse on the first qudit, but they cancel each other. Since the number of elementary operations for one column is $O(n)$, the total number of operations transforming $C$ into $I$ is $O(n^2)$.

If the entry $C_{11}$ has no inverse modulo $d$, but there is a $C_{k1}$ in the first row that does have an inverse, this entry can be switched into the first position by a permutation of two qudits and possibly the discrete Fourier transform on the first qudit. Note that it is possible that none of the entries of $C_1$ has an inverse. Indeed, since $C$ is invertible, the only restriction on one single column of $C$ is that the greatest common divisor of all its entries has an inverse. For every two entries $C_{i1}$ and $C_{n+i,1}$ or $C_{i1}$ and $C_{j1}$ from the same block, the $\gcd$ of these two can be formed in one of the two entries by recursively substracting a multiple of one row from the other following Euclid's algorithm \cite{euclid}. The other entry can then be made 0 since it is a multiple of the $\gcd$. A worst case scenario would be that all $2n$ combinations of $2n-1$ entries have a $\gcd$ that is not invertible. The procedure goes as follows
\begin{eqnarray*}
\lefteqn{\left[\begin{array}{c} C_{11}\\C_{21}\\\vdots\\C_{n1}\\\hline C_{n+1,1}\\C_{n+2,1}\\\vdots\\C_{2n,1}\end{array}\right]
\rightarrow
\left[\begin{array}{c} \gcd(C_{11},C_{n+1,1})\\\gcd(C_{21},C_{n+2,1})\\\vdots\\\gcd(C_{n1},C_{2n,1})\\\hline 0\\0\\\vdots\\0\end{array}\right]
\rightarrow}\\
&&\left[\begin{array}{c} \gcd(C_{11},\ldots,C_{2n,1})\\0\\\vdots\\0\\\hline 0\\0\\\vdots\\0\end{array}\right]
\rightarrow
\left[\begin{array}{c} 1\\0\\\vdots\\0\\\hline 0\\0\\\vdots\\0\end{array}\right].
\end{eqnarray*}
In this way, $C$ is decomposed into $O\bigl(n^2\log(d)\bigr)$ elementary operations, as the computational complexity for finding the $\gcd$ of two positive integers less than $d$ with Euclid's algorithm is $O\bigl(\log(d)\bigr)$ \cite{euclid}.

\section{Stabilizer states}\label{secS}
In this section we define stabilizer states for qudits of arbitrary dimensions. A stabilizer state is a state of an $n$-qudit system that is a simultaneous eigenvector, with eigenvalues 1, of a subgroup of $d^n$ commuting elements of the Pauli group, which is called the stabilizer ${\cal S}$ of the stabilizer state. The stabilizer state is completely determined by a generating set for ${\cal S}$. The description of such a generating set in modular arithmetic provides an efficient tool of describing the stabilizer state and its behavior under the action of a Clifford operation. Finally, we give an expansion of an arbitrary stabilizer state in the standard basis.

\subsection{Definition and description in modular arithmetic}
A stabilizer state $|\psi\rangle$ is the simultaneous eigenvector, with eigenvalues 1, of a subgroup of $d^n$ commuting elements of the Pauli group which does not contain multiples of the identity other than the identity itself. We call this subgroup the \emph{stabilizer} ${\cal S}$ of $|\psi\rangle$. A generating set for ${\cal S}$ consists of elements $\zeta^{f_k}\XZ(S_k),~k=1\ldots m$, where $S_k\in\Z_d^{2n}$ and $f_k\in\Z_{2d}$. We will assemble the vectors $S_k$ as the columns of a matrix $S\in\Z_d^{2n\times m}$ and the scalars $f_k$ in a vector $f\in\Z_{2d}^m$. We call $S$ a \emph{generator matrix} and $f$ the corresponding \emph{phase vector} that together define ${\cal S}$. The fact that the elements of ${\cal S}$ commute is reflected by $S^TPS=0\mod d$. We choose $m$ to be the minimal cardinality of a generating set of ${\cal S}$. Note that, as opposed to the situation for qubits, $m$ can be larger than $n$. It can be verified that if $m>n$, the imposed condition in Ref.~\cite{S2} for a stabilizer state to be equivalent to a graph state, is not fulfilled. If $d$ has only single prime factors, then $m=n$. If $d$ has multiple prime factors, then $n\le m\le 2n$. A simple example for $d=4$ and $n=1$ is the state $1/\sqrt{2}(|0\rangle+|2\rangle)$ with stabilizer $\{I,X^2,Z^2,X^2Z^2\}$: in this case $m=2$. We will describe below how to construct such a minimal generating set. The fact that ${\cal S}$ does not contain multiples of the identity other than the identity itself implies that the phase vector $f$ satisfies:
\begin{equation}\label{SPC}
\begin{array}{c}
\forall r\in\Z_d^m\ |\ Sr=0\mod d:\quad\bigl(f-\Vdiag(S^TUS)\bigr)^Tr+\\
r^T\bigl(\Pdiag(S^TUS)+2\Pupps(S^TUS)\bigr)r=0\mod 2d.
\end{array}
\end{equation}

The description of $\cal S$ by $S$ and $f$ is not unique, as they represent a generating set for $\cal S$. By applying an invertible linear transformation $R\in\Z_d^{m\times m}$ to the right on $S$ and transforming $f$ appropriately, another generating set $\zeta^{f_k'}\XZ(S_k')$ is formed. By repeated application of (\ref{paulimult}), one finds
\begin{equation}\label{SRC}
\begin{array}{rcl} S' & = & SR \mod d, \\
                   f' & = & R^T\bigl(f-\Vdiag(S^TUS)\bigr)+\\
                      &   & \Vdiag\Bigl(R^T\bigl(2\Pupps(S^TUS)+\\
                      &   & \qquad\quad\Pdiag(S^TUS)\bigr)R\Bigr) \mod 2d. 
\end{array}
\end{equation}
We will refer to this as a \emph{stabilizer generator matrix change}.

If $|\psi\rangle$ is operated on by a Clifford operation $Q$, defined by $C$ and $h$, then $Q|\psi\rangle$ is a new stabilizer state whose stabilizer is given by $Q{\cal S}Q^\dag$. By application of (\ref{imagepauli}), we can calculate an $S'$ and $f'$ for this stabilizer, resulting in
\begin{equation}\label{IC}
\begin{array}{rcl} S' & = & CS \mod d, \\
                   f' & = & f+S^T\bigl(h-\Vdiag(C^TUC)\bigr)+\\
                      &   & \Vdiag\Bigl(S^T\bigl(2\Pupps(C^TUC)+\\
                      &   & \qquad\quad\Pdiag(C^TUC)\bigr)S\Bigr)  \mod 2d. 
\end{array}
\end{equation}

We can construct a minimal generating set $\zeta^{f_k}\XZ(S_k),~k=1\ldots m$, for an arbitrary stabilizer ${\cal S}$, given a generating set $\zeta^{f_l'}\XZ(S_l'),~l=1\ldots m'$, for ${\cal S}$ using the Smith normal form (see Appendix~\ref{appSNF}). This can be done as follows. The $S_l'$ are assembled in the matrix $S'$. Now we compute the Smith normal form $F=KS'L$ of $S'$, with $K\in\Z_d^{2n\times 2n}$ and $L\in\Z_d^{m'\times m'}$ invertible matrices. $S'L$ is just another generator matrix of the stabilizer. From the definition of the Smith normal form it follows that $S'L$ is a generator matrix having a minimal number of nonzero columns. The rightmost $m-m'$ columns of $S'L$ that are zero (as $S'L=K^{-1}F$) can be omitted. We call this new generator matrix $S$ and $f$ is formed out of $f'$ with (\ref{SRC}). Note that no linear combination of the columns $S_k$ of $S$ is zero unless the coefficients in this linear combination are a multiple of the order of the columns, or, for $k=1\ldots m$,
\begin{equation}\label{lincombi}
\mathrm{if}~\sum_k r_k S_k=0\mod d,~\mathrm{then}~r_k S_k=0 \mod d.
\end{equation}
With this, the stabilizer phase condition (\ref{SPC}) can be simplified to, for $k=1\ldots m$:
\begin{equation}\label{SPC2}
\begin{array}{c}
\forall r_k\in\Z_d~|~r_k S_k=0\mod d:\\\quad(r_k-1)r_kS_k^TUS_k+r_kf_k=0\mod 2d.
\end{array}
\end{equation}

\subsection{Description of a stabilizer state with linear and quadratic forms}
We provide an expansion of an arbitrary stabilizer state in the standard basis for an $n$-qudit state. This is stated in the following theorem
\begin{theorem}\label{theom1}
(i) If $S\in\Z_d^{2n\times m}$ and $f\in\Z_{2d}^m$ define a stabilizer state $|\psi\rangle$ as
described above, then $S$ and $f$ can be transformed by an
configuration space transformation $|x\rangle\rightarrow|T^{-1}x\rangle$,
with $T\in\Z_d^{n\times n}$, and a stabilizer generator matrix
change $R\in\Z_d^{m\times m}$ into the form $S'$ and $f'$, with
\begin{equation}\label{eqsq1}
\begin{array}{rcl}
   S' =
   \left[\begin{array}{cc}
     T^{-1} & 0\\
     0   & T^T
   \end{array}\right] S R & = &
   \left[\begin{array}{cc}
      {\bar Q} & 0\\
      {\bar B} & {\bar {\bar B}} 
   \end{array}\right]=
   \left[\begin{array}{c}
      Q  \\
      B  
   \end{array}\right]\mod d,\\
   {f'}^T & = & \left[\begin{array}{cc} {\bar f}'^T & {\bar {\bar f}'}^T\end{array}\right]\mod 2d,
\end{array}
\end{equation}
where $Q$ is a pseudo-diagonal matrix in \emph{Smith normal form} and $Q^TB\mod d$ is symmetric. ${\bar Q}$ and ${\bar B}$ are the left square $n\times n$ parts of $Q$ and $B$.

(ii) The state $|\psi\rangle$ can be expanded in the standard basis (up to a normalization factor) as
\begin{equation}\label{psi}
 |\psi\rangle= \sum_{t\in\Z_d^n} \zeta^{t^TMt+p^Tt}\ |T({\bar Q}t+x^\ast)\rangle 
\end{equation}
where $\begin{array}[t]{rcl}
M & := & {\bar Q}{\bar B}\mod d,\\
p & := & {\bar f}'-\Vdiag(M)+2{\bar B}^Tx^\ast\mod 2d.\end{array}$\\
If we define the $n$-vector ${\bar q}$ with entries $q_k:=\left\{\begin{array}{cl}d & \mathrm{if}~Q_{kk}=0\mod d\\Q_{kk} & \mathrm{if}~Q_{kk}\not=0\mod d\end{array}\right.,~k=1\ldots n$, and the $m$-vector $q=[{\bar q}^T~\underbrace{d~\ldots~d}_{m-n}]^T$. Then $x^\ast\in G_{\bar q}:=\Z_{q_1}\times\ldots\times\Z_{q_n}$ is defined as the unique solution of 
\begin{equation}\label{x}
B^Tx=y\mod q,
\end{equation}
where $y\in G_q$ has entries $y_k:=\left\{\begin{array}{rl}-\frac{(d-q_k)B_{kk}+f_k'}{2}\mod q_k, & \mathrm{for}~k=1\ldots n \\-\frac{f_k'}{2}\mod q_k, & \mathrm{for}~k=n+1\ldots m\end{array}\right.$.\end{theorem}
Note that from the stabilizer phase condition (\ref{SPC2}) (choose $r_k:=d$), it follows that the numerators in the expressions for $y_k$ are even. An efficient way of solving (\ref{x}) can be found in Appendix~\ref{appunicity}.

A definition of the Smith normal form of a matrix $\in\Z_d^{n\times m}$ is given in Appendix~\ref{appSNF}.

{\bf Proof:} 

(i) We assume that $S$ already has a minimal number of columns $m$ as described above and we write $S$ as $\left[\begin{array}{c}S_{(1)}\\S_{(2)}\end{array}\right]$ with $S_{(1)},S_{(2)}\in\Z_d^{n\times m}$. Then we define $Q$ as the Smith normal form of $S_{(1)}$ with invertible transformation matrices $T^{-1}$ and $R$, i.e. $Q=T^{-1}S_{(1)}R$. With $B=T^TS_{(2)}R$, this yields the expression for $S'$ in (\ref{eqsq1}). According to (\ref{SRC}) and (\ref{IC}), $f$ is transformed to $f'$, yielding
\begin{eqnarray*}
f'&=& R^T\bigl(f-\Vdiag(S^TUS)\bigr)+\Vdiag\Bigl(R^T\bigl(2\Pupps(S^TUS)\\
&&+\Pdiag(S^TUS)\bigr)R\Bigr)\mod 2d.
\end{eqnarray*}
Note that $\left[\begin{array}{cc}T^{-1} & 0\\0 & T^T\end{array}\right]^TU\left[\begin{array}{cc}T^{-1} & 0\\0 & T^T\end{array}\right]=U\mod d$.
It follows directly from $S^TPS=0\mod d$ that $Q^TB$ is symmetric modulo $d$.

(ii) We show that (\ref{psi}) is a simultaneous eigenvector with eigenvalue 1 of $\zeta^{f_k}\XZ(S_k),~k=1\ldots m$. Equivalently, the state
\begin{equation}\label{psiaccent}
|\psi'\rangle:= \sum_{t\in\Z_d^n} \zeta^{t^TMt+p^Tt}\ |{\bar Q}t+x^\ast\rangle 
\end{equation}
is a simultaneous eigenvector with eigenvalue 1 of $\zeta^{f_k'}\XZ(S_k'),~k=1\ldots m$. First, note that in (\ref{psiaccent}), different values of $t$ may yield the same basis state $|{\bar Q}t+x^\ast\rangle$, since ${\bar Q}t+x^\ast\mod d$ is periodic. The coefficient of $|{\bar Q}t+x^\ast\rangle$ in (\ref{psiaccent}) displays the same periodic behavior: if ${\bar Q}t={\bar Q}t'\mod d$ then $t^TMt+p^Tt={t'}^TMt'+p^Tt'\mod 2d$. It is sufficient to check this for $t'=t+\frac{d}{q_k}E_k,~k=1\ldots n$, where $E_k$ are the standard basis vectors of $\Z_d^n$. We have
\begin{equation*}
\begin{array}{l}
(t+\frac{d}{q_k}E_k)^TM(t+\frac{d}{q_k}E_k)+p^T(t+\frac{d}{q_k}E_k)-t^TMt-p^Tt\\
=\frac{d}{q_k}\bigl((d-q_k)B_{kk}+f_k'+2B_k^Tx^\ast\bigr)=0\mod 2d,
\end{array}
\end{equation*}
for $k=1\ldots n$. Indeed, from the definition of $x^\ast$: $B_k^Tx^\ast=-\frac{(d-q_k)B_{kk}+f_k'}{2}\mod q_k,~k=1\ldots n$, it follows that $2\frac{d}{q_k}B_k^Tx^\ast=-\frac{d}{q_k}\bigl((d-q_k)B_{kk}+f_k'\bigr)\mod 2d,~k=1\ldots n$. We made use of the fact that $M={\bar Q}{\bar B}\mod d\Rightarrow M={\bar Q}{\bar B}+D\mod 2d$, where every entry of $D\mod 2d$ can be either $d$ or 0, i.e. $2D=0\mod 2d$.

Next, we check for $k=1\ldots n$ that (\ref{psiaccent}) is an eigenvector of $\zeta^{f_k'}\XZ(S_k')$ with eigenvalue 1. We have
\begin{eqnarray*}
\lefteqn{\zeta^{f_k'}\XZ\left(\left[\begin{array}{c}Q_k \\ 
                                           B_k\end{array}\right]\right)|\psi'\rangle}\\
& = & \sum_{t\in\Z_d^n} \zeta^{t^TMt+p^Tt+f_k'+2B_k^T({\bar Q}t+x^\ast)}\ |{\bar Q}t+x^\ast+Q_k\rangle \\
& = & \sum_{t\in\Z_d^n} \zeta^{(t-E_k)^TM(t-E_k)+p^T(t-E_k)+f_k'+2B_k^T({\bar Q}(t-E_k)+x^\ast)}\\
&&\qquad|{\bar Q}t+x^\ast\rangle \\
& = & \sum_{t\in\Z_d^n} \zeta^{t^TMt+p^Tt}\ |{\bar Q}t+x^\ast\rangle = |\psi'\rangle.\\
\end{eqnarray*}

Finally, $\zeta^{f_k'}\XZ(S_k')$ acting on the left of (\ref{psiaccent}) yields, for $k=n+1\ldots m$,
\begin{eqnarray*}
\lefteqn{\zeta^{f_{k}'}\XZ\left(\left[\begin{array}{c}0 \\ 
                                             B_k\end{array}\right]\right)|\psi'\rangle}\\
& = & \sum_{t\in\Z_d^n} \zeta^{t^TMt+p^Tt+f_k'+2B_k^T({\bar Q}t+x^\ast)}\ |{\bar Q}t+x^\ast\rangle \\
& = & \sum_{t\in\Z_d^n} \zeta^{t^TMt+p^Tt}\ |{\bar Q}t+x^\ast\rangle = |\psi'\rangle.\\
\end{eqnarray*}
In Appendix~\ref{appunicity} we prove that eq.~(\ref{x}) has a unique solution $x^\ast\in G_{\bar q}$. \hfill$\square$

It is possible to remove all identical terms in the summation of expression~(\ref{psi}) as follows. We define $r$ as the number of nonzero diagonal elements of $Q$. We denote the upper left $r\times r$-part of a matrix $A$ as $A_{(r)}$, the upper $r$-part of a vector $a$ as $a_{(r)}$ and the part of $a$ below $a_{(r)}$ as ${\bar a}_{(r)}$. Then (\ref{psi}) is equivalent to
\begin{equation*}
|\psi\rangle= \sqrt{\frac{\prod_{i=1}^{r}q_i}{d^r}}\sum_{t\in G_\ast} \zeta^{t^TM_{(r)}t+p_{(r)}^Tt}\ |T\left[\begin{array}{c}Q_{(r)}t+x_{(r)}^\ast\\{\bar x}_{(r)}^\ast\end{array}\right]\rangle,
\end{equation*}
where $G_\ast:=\Z_{\frac{d}{q_1}}\times\ldots\times\Z_{\frac{d}{q_r}}$. Note that the normalizing factor is just the inverse of the square root of the number of terms in the summation, as each basis state is orthogonal to the others and occurs only once.
Finally, it is interesting to mention that, for an arbitray $S$ and $f$ defining a stabilizer state $|\psi\rangle$, we have (up to a normalization factor)
\begin{equation*}
|\psi\rangle= \sum_{t\in\Z_d^m} \zeta^{t^TMt+p^Tt}\ |S_{(1)}t+x'\rangle,
\end{equation*}
where $S=\left[\begin{array}{c}S_{(1)}\\S_{(2)}\end{array}\right]$, $M=S_{(1)}^TS_{(2)}\mod d$, $p=f-\Vdiag(M)+2S_{(2)}^Tx'\mod 2d$ and $x'=Tx^\ast\mod d$, where $T$ and $x^\ast$ are the same as in (\ref{psi}). Yet, this formula has two disadvantages: first, to find $x'$, we still have to calculate the Smith normal form of $S_{(1)}$ and second, in (\ref{psi}) it is clearer which basis states have nonzero coefficients.

\section{Conclusion}
We have shown that for the Pauli group, the Clifford group and stabilizer states, straightforward extensions in Hilbert spaces of arbitrary dimensions can be compactly described with matrices over $\Z_d$. We have given a way of efficiently decomposing an $n$-qudit Clifford operation in $O(n^2)$ one and two-qudit operations. With these tools in modular arithmetic, we provide an expansion of an arbitrary stabilizer state of $n$ qudits in the standard basis.

\appendix
\section{The Smith normal form}\label{appSNF}
The Smith normal form is a canonical diagonal form for equivalence of matrices over a principal ideal ring $R$. In this paper we consider matrices over $\Z_d$. For any $A\in\Z_d^{n\times m}$ there exist invertible matrices $K\in\Z_d^{n\times n}$ and $L\in\Z_d^{m\times m}$ such that
\begin{equation*}
F=KAL=\left[\begin{array}{cccccc}
f_1 & & & & & \\ & \ddots & & & & \\ & & f_r & & & \\ & & & 0 & & \\ & & & & \ddots & \\ & & & & & 0
\end{array}\right]\mod d
\end{equation*}
with each $f_i$ a nonzero and with $f_i|f_{i+1}$ for $1\le i\le r-1$. The $f_i$ are unique up to units. Uniqueness of $F$ can be ensured by specifying that each $f_i$ should be a positive divisor of $d$ in $\Z$. There exist fast algorithms for computing the Smith normal form \cite{stor}.

\section{A unique solution of (\ref{x})}\label{appunicity}
Here we prove that eq.~(\ref{x}) $B^Tx=y\mod q$ has a unique solution $x^\ast\in G_{\bar q}:=\Z_{q_1}\times\ldots\times\Z_{q_n}$. We rewrite (\ref{x}) as the following system of equations:
\begin{equation}\label{system}
\sum_{i=1}^{n}B_{ij}x_i=y_j\mod q_j,~j=1\ldots m.
\end{equation}
If, for fixed $j$, the $B_{ij}$ and $q_j$ have a common factor, then also $y_j$ must be a multiple of this factor, otherwise there is no solution. Define $g_j:=\gcd(B_{1j},\ldots,B_{nj},q_j)$ and $r_j:=d/g_j$. Note that $r_j$ is the order of $S_j$. A necessary condition for solvability of (\ref{system}) is or $r_jy_j=0\mod d$, for every $k=1\ldots m$. We show that this condition holds. We have $r_jS_j=0\mod d$. From the stabilizer phase condition~(\ref{SPC2}), it follows that
\begin{equation*}
\begin{array}{rclc}
(r_j-1)r_jq_jB_{jj}+r_jf_j' & = & 0\mod 2d, & 1\le j\le n,\\
r_jf_j' & = & 0\mod 2d, & j>n,
\end{array}
\end{equation*}
and by definition of $y$, consequently $r_jy_j=0\mod d,~j=1\ldots m$.

An equivalent system to (\ref{system}) is now
\begin{equation*}
\sum_{i=1}^{n}\frac{B_{ij}}{g_j}x_i=\frac{y_j}{g_j}\mod \frac{q_j}{g_j},~j=1\ldots m.
\end{equation*} 
We define the map $b:x=[x_1~\ldots~x_n]^T\rightarrow b(x)=\left[\sum_{i=1}^n\frac{B_{i1}}{g_1}x_i|\ldots|\sum_{i=1}^n\frac{B_{im}}{g_m}x_i\right]^T\mod\left[\frac{q_1}{g_1}\ldots\frac{q_m}{g_m}\right]^T$, which is a homomorphism from the group of vectors of length $n$ with entries $x_i$ modulo $q_i,~i=1\ldots n$ to the group of vectors of length $m$ with entries $y_j'$ modulo $q_j/g_j,~j=1\ldots m$. (\ref{system}) has a unique solution if $b$ is an isomorphism. We prove this by showing that the number of elements in both groups are the same and that only 0 is in the kernel. It follows from (\ref{lincombi}) and the fact that, by definition, the columns of $S$ generate a set of $d^n$ elements, that the product of the orders of the columns of $S$ is equal to $d^n$, or $\prod_{j=1}^{m}r_j=d^n$. Therefore,
\begin{equation*}
\prod_{j=1}^{m}\frac{q_j}{g_j} = \frac{d^{m-n}}{\prod_{j=1}^{m}g_j}\prod_{i=1}^{m}q_i = \prod_{i=1}^{m}q_i
\end{equation*}
thus the number of elements of both groups are the same. Next we show that $B^Tx=0\mod q$ if and only if $x=0\mod{\bar q}$. We rewrite this as
\begin{equation}\label{isomo}
\begin{array}{cl}
\forall x\in\Z_d^n: & \left(\exists v\in\Z_d^n: B^Tx=Q^Tv\mod d\right)\\ 
& \iff \left(\exists x'\in\Z_d^m: x=Qx'\mod d\right).
\end{array}
\end{equation}

{\bf Proof:}

$\Leftarrow$) $Q^TB$ is symmetric modulo $d$. We therefore have $B^Tx=B^TQx'=Q^TBx'\mod d$, so $v=Bx'\mod d$.

$\Rightarrow$) We show that the number of $x\in\Z_d^n$ satisfying the lhs of (\ref{isomo}) is equal to the number of $x$ satisfying the rhs. The number of elements generated by the columns of a matrix is equal to the product of the orders of the diagonal elements of its Smith normal form. Therefore the columns of $S^T$, like the columns of $S$, also generate $d^n$ elements. Consequently, the mapping $s: a\in\Z_d^{2n}\rightarrow S^Ta\in\Z_d^{m}$ is a homomorphism from $\Z_d^{2n}$ to a group $Y\subset\Z_d^{m}$, with $|Y|=d^n$. The kernel in $\Z_d^{2n}$ of $s$ contains $|\Z_d^{2n}|/|Y|=d^n$ elements. Equivalently, with $a^T=[v^T~w^T]$, $s$ is a homomorphism from $\Z_d^{n}\times\Z_d^{n}$ to $Y$:
\begin{equation*}
s(\left[\begin{array}{c}v\\w\end{array}\right])=S^T\left[\begin{array}{c}v\\w\end{array}\right]=\left[\begin{array}{cc}Q^T & B^T\end{array}\right]\left[\begin{array}{c}v\\w\end{array}\right]=Q^Tv+B^Tw.
\end{equation*}
There are exactly $d^n$ different pairs $(v,w)$ that satisfy $Q^Tv+B^Tw=0\mod d$. Replacing $w$ by $-x$, we have exactly $d^n$ pairs $(x,v)$ satisfying $B^Tx=Q^Tv\mod d$. Fixing such an $x$, we have a total of $\prod_{i=1}^{n}q_i$ different $v$ for which, together with $x$, the equality still holds (this is because we can add an arbitrary multiple of $d/q_i$ to $v_i$). Therefore, the total number of $x$ for which a $v$ exists such that $B^Tx=Q^Tv\mod d$, is equal to $d^n/\prod_{i=1}^{n}q_i$. This is equal to the number of $x$ that can be written as $x=Qx'$. \hfill$\square$

Next, we describe a method for easily finding the solution $x^\ast$ of (\ref{x}). We define a diagonal matrix $Z\in\Z_d^{m\times m}$ with diagonal entries equal to $d/q_k,~k=1\ldots m$. (\ref{x}) is equivalent to the equation $ZB^Tx=Zy\mod d$.
We calculate the Smith normal form $F=KZB^TL\mod d$. Defining $x':=L^{-1}x\mod d$ and $y':=KZy\mod d$, we have the following equation $Fx'=y'\mod d$, for which a solution ${x^\ast}'\in\Z_d^n$ can be easily found (note that this solution is most likely not unique). We then find $x^\ast=L{x^\ast}'\mod {\bar q}$.

\section{Simplifications for odd $d$}\label{appsimp}
In this section we consider the special case of odd $d$. Most of the formulas in this paper can be simplified for odd $d$. We will only give an overview and omit the derivations, as they are completely analogous to the general case. If $d$ is odd, then 2 has an inverse in $\Z_d$, equal to $\frac{d+1}{2}$, which we will denote by $2^{-1}$.

For odd $d$, we can use a restricted definition for the Pauli group: it contains all $d^{2n}$ tensor products (\ref{XZ(a)}) with an additional complex phase factor $\omega^\delta$ (instead of a power of $\zeta$). Eq. (\ref{paulimult}) becomes
\begin{equation}
\omega^\delta\XZ(a)\omega^\epsilon\XZ(b)=\omega^{\delta+\epsilon+a^TUb}\XZ(a+b).
\end{equation}
The order of an arbitrary element of this newly defined Pauli group is never equal to $2d$. In the same way as for the general case, we find the image $\omega^\epsilon\XZ(b)$ of $\omega^\delta\XZ(a)$ under conjugation by a Clifford operation, which is now defined by $C$ and $g=\frac{h}{2}$:
\begin{equation}
\begin{array}{rcl} b & = & Ca \mod d, \\
            \epsilon & = & \delta+\left(g-2^{-1}\Vdiag(C^TUC)\right)^Ta+\\
                     &   & 2^{-1}a^T(C^TUC-U)a \mod d. \end{array}
\end{equation}
Note that, contrary to the general case, $g$ is a vector in $\Z_d^{2n}$. There is no longer a restriction on $g$. Indeed, from (\ref{condh}), it follows that $h$ in the general setting is always even for odd $d$. Symplecticity of $C$ is of course still required. The product of two Clifford operations $Q''=Q'Q$ corresponds to $C''$ and $g''$, where
\begin{equation}
\begin{array}{rcl} C'' & = & C'C \mod d, \\
                   g'' & = & g+C^Tg'+2^{-1}\Bigl(\Vdiag\bigl(C^T({C'}^TUC'-U)C\bigr)-\\
                       &   & C^T\Vdiag({C'}^TUC')\Bigr) \mod d. \end{array}
\end{equation}
The inverse $Q^\dag$ of a Clifford operation $Q$ defined by $C$ and $g$ is defined by $C'$ and $g'$, where
\begin{equation}
\begin{array}{rcl} C' & = & C^{-1} = -PC^TP \mod d, \\
                   g' & = & -C^{-T}g+2^{-1}\Bigl(C^{-T}\Vdiag(C^TUC)+\\
                      &   & \Vdiag(C^{-T}UC^{-1})\Bigr) \mod d. \end{array}
\end{equation}

The definition of a stabilizer state remains the same except for the fact that now the stabilizer is a subgroup of the restricted Pauli group. Note that for odd $d$, no subgroup of the general Pauli group can be found that fulfills all stabilizer conditions but is not a subgroup of the restricted Pauli group. Thus, nothing is lost by restricting the definition of the Pauli group for odd $d$. A generating set for the stabilizer ${\cal S}$ consists of elements $\omega^{b_k}\XZ(S_k),~k=1\ldots m$, where $S_k\in\Z_d^{2n}$ and $b_k\in\Z_d^m$. Analogously to the definition of $g$, $b$ is equal to half the value of $f$ in the general setting (as it is the exponent of $\omega$ instead of $\zeta$). The stabilizer phase condition (\ref{SPC}) on $b$ simplifies to: 
\begin{equation}
\begin{array}{c}
\forall r\in\Z_d^m\ |\ Sr=0\mod d:\\
\bigl(2b-\Vdiag(S^TUS)\bigr)^Tr+r^T(S^TUS)r=0\mod d.
\end{array}
\end{equation}
A stabilizer generator matrix change, by applying an invertible linear transformation $R\in\Z_d^{m\times m}$ to the right on $S$, results in
\begin{equation}
\begin{array}{rcl} S' & = & SR \mod d, \\
                    b' & = & R^T\bigl(b-2^{-1}\Vdiag(S^TUS)\bigr)+\\
                       &   & 2^{-1}\Vdiag\bigl(R^TS^TUSR\bigr) \mod d. 
\end{array}
\end{equation}
A stabilizer state defined by $S$ and $b$, operated on by a Clifford operation defined by $C$ and $g$, is a new stabilizer state defined by
\begin{equation}
\begin{array}{rcl} S' & = & CS \mod d, \\
                    b' & = & b+S^T\bigl(g-2^{-1}\Vdiag(C^TUC)\bigr)+\\
                       &   & 2^{-1}\Vdiag\bigl(S^T(C^TUC-U)S\bigr) \mod d. 
\end{array}
\end{equation}
It is not hard to verify that part \emph{(ii)} of Theorem~\ref{theom1} simplifies to
\begin{equation}
 |\psi\rangle= \sum_{t\in\Z_d^n} \omega^{t^TMt+p^Tt}\ |T({\bar Q}t+x^\ast)\rangle 
\end{equation}
where $\begin{array}[t]{rcl}
M & := & 2^{-1}{\bar Q}{\bar B}\mod d,\\
p & := & {\bar b}'-\Vdiag(M)+{\bar B}^Tx^\ast\mod d.\end{array}$\\
In this setting, $x^\ast\in G_{\bar q}:=\Z_{q_1}\times\ldots\times\Z_{q_n}$ is defined as the unique solution of $B^Tx=-b'\mod q$. For calculating $x^\ast$ we refer to Appendix~\ref{appunicity}.

\begin{acknowledgments}
We thank Maarten Van den Nest for useful comments.
Dr. Bart De Moor is a full professor at the Katholieke Universiteit Leuven, Belgium. Research supported by: Research Council KUL: GOA-Mefisto~666, GOA AMBioRICS, several PhD/postdoc \& fellow grants; Flemish Government: FWO: PhD/postdoc grants, projects, G.0240.99 (multilinear algebra), G.0407.02 (support vector machines), G.0197.02 (power islands), G.0141.03 (Identification and cryptography), G.0491.03 (control for intensive care glycemia), G.0120.03 (QIT), G.0452.04 (new quantum algorithms), G.0499.04 (Robust SVM), research communities (ICCoS, ANMMM, MLDM); AWI: Bil. Int. Collaboration Hungary/Poland; IWT: PhD Grants, GBOU (McKnow); Belgian Federal Science Policy Office: IUAP P5/22 (`Dynamical Systems and Control: Computation, Identification and Modelling', 2002-2006) ; PODO-II (CP/40: TMS and Sustainability); EU: FP5-Quprodis; ERNSI; Eureka 2063-IMPACT; Eureka 2419-FliTE; Contract Research/agreements: ISMC/IPCOS, Data4s, TML, Elia, LMS, Mastercard.
\end{acknowledgments}

\end{document}